\documentclass[letterpaper, 10 pt, journal, twoside]{IEEEtran}
\let\authorrefmark\IEEEauthorrefmark

\usepackage{style}
\usepackage{booktabs}

\title{\LARGE Regret-Optimal Defense Against Stealthy Adversaries:\\A System Level Approach}

\author{Hiroyasu Tsukamoto\authorrefmark{1}\authorrefmark{3}, Joudi Hajar\authorrefmark{2}, Soon-Jo Chung\authorrefmark{2}\authorrefmark{3}, and Fred Y. Hadaegh\authorrefmark{2}\authorrefmark{3}
\thanks{\authorrefmark{1} Department of Aerospace Engineering, University of Illinois at Urbana-Champaign, Urbana, IL, {\tt\footnotesize \href{mailto:hiroyasu@illinois.edu}{hiroyasu@illinois.edu}}.}
\thanks{\authorrefmark{2} Division of Engineering and Applied Science, Caltech, Pasadena, CA, {\tt\footnotesize\{\href{mailto:jhajar@caltech.edu}{jhajar}, \href{mailto:sjchung@caltech.edu}{sjchung}, \href{mailto:hadaegh@caltech.edu}{hadaegh}\}@caltech.edu}.}
\thanks{\authorrefmark{3} Jet Propulsion Laboratory, Pasadena, CA. This project is funded by the Technology Innovation Institution (TII) with a contract with Caltech. The first author is funded by the University of Illinois at Urbana-Champaign. Part of the research was carried out at the Jet Propulsion Laboratory, California Institute of Technology, under a contract with the National Aeronautics and Space Administration.}}

\begin{document}
\maketitle

\begin{abstract}
Modern control designs in robotics, aerospace, and cyber-physical systems rely heavily on real-world data obtained through system outputs. However, these outputs can be compromised by system faults and malicious attacks, distorting critical system information needed for secure and reliable operation. In this paper, we introduce a novel regret-optimal control framework for designing controllers that make a linear system robust against stealthy attacks, including both sensor and actuator attacks. Specifically, we present (a) a convex optimization-based system metric to quantify the regret under the worst-case stealthy attack (the difference between actual performance and optimal performance with hindsight of the attack), which adapts and improves upon the $\mathcal{H}_2$ and $\mathcal{H}_{\infty}$ norms in the presence of stealthy adversaries, (b) an optimization problem for minimizing the regret of (a) in system-level parameterization, enabling localized and distributed implementation in large-scale systems, and (c) a rank-constrained optimization problem equivalent to the optimization of (b), which can be solved using convex rank minimization methods. We also present numerical simulations that demonstrate the effectiveness of our proposed framework. 
\end{abstract}
\section{Introduction}
\label{sec_introduction}
Recent advances in autonomous control, often enhanced by machine learning, highlight the critical role of real-world data in building resilient autonomous systems and their decision-making processes. When using such data in robotics, aerospace, and cyber-physical systems, a critical question arises: ``What if the system outputs are compromised and fail to capture important real-world data, potentially indicating a system anomaly?''

This paper addresses this issue by examining discrete-time linear time-varying systems affected by stealthy actuator and sensor disturbances. We define a disturbance as stealthy when the difference between the system outputs, with and without disturbances, stays below a certain threshold. These disturbances can lead to the loss of crucial adversarial information in real-world data, occurring in scenarios such as (1) malicious sensor and actuator attacks, (2) system faults like actuator or sensor failures, and (3) disturbances or uncertainties masked by the system's output. Developing a framework to detect and optimally mitigate these stealthy disturbances is crucial for tackling problems in cybersecurity, Fault Detection, Identification, and Recovery (FDIR), thereby ensuring reliable autonomy in robotics and aerospace systems.

\subsubsection*{Contributions} We propose a novel rank minimization framework to derive the regret-optimal system-level parameterization, specifically designed to enhance resilience against stealthy actuator and sensor disturbances. The regret in the presence of adversarial disturbances is defined as $\text{Regret} = J_t - J_c$, where $J_t$ represents the performance using only causal information, and $J_c$ represents the performance with non-causal information~\cite{regret_original,regret_extensive,regret_optimal_control}, \ie{},
\begin{align}
    J_t =&~\textit{true performance with access only to past attacks} \nonumber \\
    J_c =&~\textit{clairvoyant performance with access to past, present,} \nonumber\\
    &~\textit{and future attacks}. \nonumber
\end{align}
Note that the resultant regret-optimal controller does not require any knowledge of future attacks, although the clairvoyant $J_c$ depends on them: it measures the control performance based on what it could have done if it had known about future attacks. Our method further leverages the System-Level Synthesis (SLS) parameterization introduced in~\cite{sls_tutorial,sls_output_feedback_control,sls_ARC,slsregret}, which provides a necessary and sufficient characterization of the achievable system response for the entire closed-loop system. It reframes the conventional control synthesis problem as designing the entire closed-loop system, enabling the integration of a wide range of system-level constraints, such as distributed constraints, localizability, and robustness, within a convex formulation~\cite{sls_ARC}. We achieve the following contributions by combining the approaches of regret optimality and SLS in systems with stealthy adversaries.

\begin{enumerate}[label={\color{uiucblue}{(\alph*)}}]
\item We introduce a new convex optimization-based system metric that quantifies regret under the worst-case stealthy attack. This metric adjusts between the average and robust performances of $\mathcal{H}_2$ and $\mathcal{H}_{\infty}$ norms in the presence of stealthy adversaries.\label{item_a}
\item We formulate an optimization problem for minimizing the regret of~\ref{item_a} based on system-level parameterization, enabling scalable implementation in distributed and localized control systems.\label{item_b}
\item We reformulate the optimization problem of~\ref{item_b} as a rank-constrained problem, featuring a convex objective with convex constraints and rank restrictions.
\end{enumerate}
The optimization problem in the third contribution is convex except for the rank constraints, which can be addressed using established convex rank minimization techniques~\cite{BoydRankRelaxation,rank_constrained_kyotoU,rank_constrained_boyd,rank_constrained_microsoft,rank_constrained_pourdue_ran}.
\subsubsection*{Related Work}
Regret-optimal control~\cite{regret_original,regret_extensive,regret_optimal_control} is an adaptive framework that minimizes the performance gap, or regret, between a causal controller and an ideal clairvoyant (non-causal) controller. Building on concepts from online learning and adaptive control~\cite{adaptive_regret1,learning_regret1,learning_regret2,learning_regret3,regretboffi,NEURIPS2020_155fa095,NEURIPS2019_6d1e481b}, it dynamically adjusts control inputs based on real-time observations, aiming to minimize worst-case regret. While inspired by the minimax approach of $\mathcal{H}_{\infty}$ control~\cite{Basar1995}, regret-optimal control is less conservative due to its consideration of clairvoyant information, striking a balance between the robust performance of $\mathcal{H}_{\infty}$ and the average performance of $\mathcal{H}_2$~\cite{regret_dissertation,regret_partialobs,slsregret}.

Recent work~\cite{slsregret} shows that dynamic regret-optimal controllers can be synthesized using system-level parameterization~\cite{sls_tutorial,sls_output_feedback_control,sls_ARC}. This method enables the design of the entire closed-loop system response, expanding the range of controllers that can be implemented through convex optimization.

This paper extends these methods to handle more challenging scenarios involving stealthy adversaries (see~\cite{jing2,jing3} and~\cite{security_metric_book,outputL2gain,secure_control,zda_discrete,attack_bullo} for details on stealthy adversaries). In particular, the security metric we derive generalizes the output-to-output $\ell_2$ gain~\cite{security_metric_book,outputL2gain} by incorporating regret-optimal control and system-level parameterization. This approach retains the adaptive and robust defense capabilities, making it well-suited for large-scale linear systems facing stealthy attacks. In Sec.\ref{sec_extension}, we propose ways to enhance this approach by incorporating the useful properties of system-level parameterization and regret-optimal control, such as sparsity and localizability, within an online learning framework. This fact demonstrates the strong potential for data-driven applications\cite{data_driven_sls,slsmpc,data_sls_mpc,sls_data_predictive}.


\section{Preliminaries }
\label{sec_preliminaries}
For $A \in \mathbb{R}^{n \times n}$, we use $A \succ 0$, $A \succeq 0$, $A \prec 0$, and $A \preceq 0$ to denote its definiteness. For $x \in \mathbb{R}^n$, $\|x\|$ represents the Euclidean norm. We also define $\blkdiag$, $\rank$, and $\matvec$ as the block diagonal, matrix rank, and vectorization functions, respectively. Lastly, $\mathbb{O}$ and $\mathbb{I}$ represent the zero and identity matrices, respectively.
In this paper, we study the following discrete-time, time-varying, networked linear control system:
\begin{subequations}
\label{eq_system}
\begin{align}
    x_{k+1} &= A_kx_k +B_{u,k}u(y_k,\nu_k)+B_{a,k}a_k\label{eq_continuous_initialx}\textcolor{white}{\eqref{eq_continuous_initialx}}\\
    y_k &= C_{y,k}x_k+D_{ya,k}a_k\label{eq_continuous_initialy}\textcolor{white}{\eqref{eq_continuous_initialy}}\\
    z_k &= C_{z,k}x_k+D_{zu,k}u_k\label{eq_continuous_initialz}\textcolor{white}{\eqref{eq_continuous_initialz}}
\end{align}
\end{subequations}
where $\nu_k$ represents the network topology (e.g., adjacency matrix), $x_k$ is the system state, $u$ is the control input, $a_k$ is the attack input where $B_{a,k}a_k$ represents the actuator attack and $D_{ya,k}a_k$ accounts for the sensor attack, $y_k$ is the system measurement, and $z_k$ is the regulated output used to evaluate control performance. 

\subsubsection*{Operator Formalism}As in~\cite{sls_ARC}, we use the following notations for the system signals: $\mathbf{x} = [x_0^{\top},\cdots,x_{T}^{\top}]^{\top}$, $\mathbf{u} = [u_0^{\top},\cdots,u_{T}^{\top}]^{\top}$, $\mathbf{a} = [x_0^{\top},a_0^{\top}\cdots,a_{T-1}^{\top}]^{\top}$, $\mathbf{y} = [y_0^{\top},\cdots,y_{T}^{\top}]^{\top}$, and $\mathbf{z} = [z_0^{\top},\cdots,z_{T}^{\top}]^{\top}$, where $u_k = u(y_k,\nu_k)$ and $T$ is the time horizon. We also use the following notations for the system matrices: $
    \mathcal{A} = \blkdiag(A_0,\cdots,A_{T-1},\mathbb{O})$, 
    $\mathcal{B}_u = \blkdiag(B_{u,0},\cdots,B_{u,T-1},\mathbb{O})$,
    $\mathcal{B}_a = \blkdiag(\mathbb{I},B_{a,0},\cdots,\allowbreak B_{a,T-1})$,
    $\mathcal{C}_{{p}} = \blkdiag(C_{{p},0},\cdots,C_{{p},T})$, and
    $\mathcal{D}_{{p}{q}} = \blkdiag(\allowbreak D_{{p}{q},0},\cdots,D_{{p}{q},T})$,
where ${q} \in \{u,a\}$, ${p} \in \{y,z\}$, and $\blkdiag$ denotes the block diagonalization. Then \eqref{eq_system} can be written as
\begin{subequations}
\label{eq_system_bf}
\begin{align}
    \mathbf{x} &= \mathcal{Z}\mathcal{A}\mathbf{x}+\mathcal{Z}\mathcal{B}_{u}\mathbf{u}+\mathcal{B}_{a}\mathbf{a}\label{eq_system_bf_x}\textcolor{white}{\eqref{eq_system_bf_x}} \\
    \mathbf{y} &= \mathcal{C}_{y}\mathbf{x}+\mathcal{D}_{ya}\mathbf{a}\label{eq_system_bf_y}\textcolor{white}{\eqref{eq_system_bf_y}} \\
    \mathbf{z} &= \mathcal{C}_{z}\mathbf{x}+\mathcal{D}_{zu}\mathbf{u}\label{eq_system_bf_z}\textcolor{white}{\eqref{eq_system_bf_z}}
\end{align}    
\end{subequations}
where $\mathcal{Z}$ is the block-downshift operator, \ie{}, a matrix with identity matrices along its first block sub-diagonal and zeros elsewhere.

\subsubsection*{Control}
We adopt the following output feedback control, parameterized by the network topology $\bm{\nu} = (\nu_0,\cdots,\nu_T)$: \begin{align} \label{eq_output_feedback} \mathbf{u} = \mathbf{K}(\bm{\nu})\mathbf{y} \end{align} This framework includes state estimation-based feedback control~\cite{sls_output_feedback_control}, defined by $\xi_{k+1} = A_{e,k}\xi_k + B_{e,k}y_k$ and $u_k = C_{e,k}\xi_k + D_{e,k}y_k$, where $\xi_k$ represents the internal state of the controller.
\subsection{System Level Parameterization}
\label{sec_sls}
Applying the feedback control from~\eqref{eq_output_feedback}, we get the following system dynamics:
\begin{subequations}
\label{eq_system_with_feedback}
\begin{align}
    \label{eq_system_with_feedback1}
    \begin{bmatrix}
        \mathbf{x} \\
        \mathbf{u}
    \end{bmatrix}
    &=
    \begin{bmatrix}
        \mathbf{R}\mathcal{B}_a +\mathbf{N}\mathcal{D}_{ya} \\
        \mathbf{M}\mathcal{B}_a+  \mathbf{L}\mathcal{D}_{ya}
    \end{bmatrix}\mathbf{a} = 
    \begin{bmatrix}
        \mathbf{\Phi}_x  \\
        \mathbf{\Phi}_u
    \end{bmatrix}
    \mathbf{a}\textcolor{white}{\eqref{eq_system_with_feedback1}} \\
    \mathbf{R} &= (\mathbb{I}-\mathcal{Z}\mathcal{A}-\mathcal{Z}\mathcal{B}_u\mathbf{K}\mathcal{C}_y)^{-1}\label{eq_system_with_feedback2}\textcolor{white}{\eqref{eq_system_with_feedback2}} \\
    \mathbf{N} &= \mathbf{R}\mathcal{Z}\mathcal{B}_{u}\mathbf{K} \label{eq_system_with_feedback3} \textcolor{white}{\eqref{eq_system_with_feedback3}}\\
    \mathbf{M} &=\mathbf{K}\mathcal{C}_y\mathbf{R}\label{eq_system_with_feedback4}\textcolor{white}{\eqref{eq_system_with_feedback4}} \\
    \mathbf{L} &= \mathbf{K}+\mathbf{K}\mathcal{C}_y\mathbf{R}\mathcal{Z}\mathcal{B}_u\mathbf{K} \label{eq_system_with_feedback5}\textcolor{white}{\eqref{eq_system_with_feedback5}}
\end{align}
\end{subequations}
These equations satisfy the following affine relations: 
\begin{subequations}
\label{eq_sls_conditions}
\begin{align}
    \label{eq_sls1}
    &\begin{bmatrix}
        \mathbb{I}-\mathcal{Z}\mathcal{A} & -\mathcal{Z}\mathcal{B}_u
    \end{bmatrix} 
    \begin{bmatrix}
        \mathbf{R} & \mathbf{N} \\
        \mathbf{M} & \mathbf{L}
    \end{bmatrix}
    =
    \begin{bmatrix}
        \mathbb{I} & \mathbb{O}
    \end{bmatrix} \\
    \label{eq_sls2}
    &\begin{bmatrix}
        \mathbf{R} & \mathbf{N} \\
        \mathbf{M} & \mathbf{L}
    \end{bmatrix}
    \begin{bmatrix}
        \mathbb{I}-\mathcal{Z}\mathcal{A} \\
        -\mathcal{C}_y
    \end{bmatrix}
    =
    \begin{bmatrix}
        \mathbb{I} \\
        \mathbb{O}
    \end{bmatrix}.
\end{align}
\end{subequations}
ensuring both necessary and \textit{sufficient} conditions for the existence of an output feedback controller. This formulation allows us to leverage the system-level parameterization of~\eqref{eq_system_with_feedback} for control design~\cite{sls_output_feedback_control, sls_tutorial, sls_ARC}.
\begin{lemma}
\label{lemma_sls}
The affine subspace defined by~\eqref{eq_sls_conditions} parameterizes all possible system responses~\eqref{eq_system_with_feedback} achievable by the output feedback control~\eqref{eq_output_feedback}. Furthermore, for any matrices $(\mathbf{R},\mathbf{M},\mathbf{N},\mathbf{L})$ satisfying~\eqref{eq_sls_conditions}, the feedback control~\eqref{eq_output_feedback} with $\mathbf{K} = \mathbf
{L}-\mathbf{M}\mathbf{R}^{-1}\mathbf{N}$ achieves the desired response~\eqref{eq_system_with_feedback}.
\end{lemma}
\begin{proof}
The first part can be shown just by verifying~\eqref{eq_sls_conditions} for $(\mathbf{R},\mathbf{M},\mathbf{N},\mathbf{L})$ of~\eqref{eq_system_with_feedback}. The second part follows from substituting $\mathbf{K} = \mathbf
{L}-\mathbf{M}\mathbf{R}^{-1}\mathbf{N}$ into~\eqref{eq_system_bf} with the conditions of~\eqref{eq_sls_conditions}, which results in~\eqref{eq_system_with_feedback}. See~\cite{sls_output_feedback_control,sls_tutorial,sls_ARC} for details.
\end{proof}

Thanks to Lemma~\ref{lemma_sls}, we can directly work with the system level parameterization~\eqref{eq_system_with_feedback} and the affine conditions to look for a controller~\eqref{eq_output_feedback} with a desired property. It also enables localized and distributed control implementation, which is suitable for large-scale networked control systems in this paper.
\subsection{Regret-Optimal Control}
\label{sec_pre_regret}
Let $J(\mathbf{u},\mathbf{a})$ be some performance objective to be minimized for the control and attack input defined in~\eqref{eq_system_bf}. The dynamic regret of $J$, which measures the gap between the actual optimal performance and the optimal performance in hindsight, is defined as:
\begin{subequations}
\label{eq_Jdefinition}
\begin{align}
    &\regret_{J}(\mathbf{u},\mathbf{a}) = J(\mathbf{u},\mathbf{a})-J^*(\mathbf{a}) \label{eq_Jdefinition1}\textcolor{white}{\eqref{eq_Jdefinition1}} \\
    &J^*(\mathbf{a}) = J(\mathbf{u_{nc}}(\mathbf{a}),\mathbf{a}) \label{eq_Jdefinition2}\textcolor{white}{\eqref{eq_Jdefinition2}}
\end{align}
\end{subequations}
where $\mathbf{u}_{nc}$ represents the optimal non-causal control policy with access to past and future attacks $\mathbf{a}$.

Regret-optimal control seeks to minimize the worst-case regret, denoted $\regret_{J}^{*}(\mathbf{u})$~\cite{regret_original,regret_extensive,regret_optimal_control}, as:
\begin{subequations}
\label{eq_regret_policy}
\begin{align}
    &\mathbf{u}^{*} = \mathrm{arg}\min_{\mathbf{u}}\regret_{J}^*(\mathbf{u}) \label{eq_regret_policy1}\textcolor{white}{\eqref{eq_regret_policy1}} \\
    &\regret_{J}^*(\mathbf{u}) = \max_{\|\mathbf{a}\| \leq 1}\regret_{J}(\mathbf{u},\mathbf{a}) \label{eq_regret_policy2}\textcolor{white}{\eqref{eq_regret_policy2}}
\end{align}
\end{subequations}
This approach balances two extremes: the robustness of $\mathcal{H}_{\infty}$, which may be overly conservative by focusing on worst-case scenarios, and the efficiency of $\mathcal{H}_2$, which can be too optimistic under adversarial disturbances~\cite{regret_original,regret_optimal_control}. It is also applicable in the output feedback context considered in this work, as discussed in~\cite{regret_measurement,regret_partialobs,regret_dissertation}. Additionally, recent results~\cite{slsregret} demonstrate the compatibility of regret-optimal control with the system-level parameterization in Sec.~\ref{sec_sls}.
\section{Problem Formulation}
We define a class of stealthy attacks called $\alpha$-stealthy attacks, consistent with the approach in~\cite{outputL2gain}, which broadens the concept of stealthiness. When $\alpha = 0$, it describes conventional stealthy attacks where the detector’s output is always zero, such as zero dynamics attacks~\cite{zda_discrete,secure_control}. For $\alpha > 0, $ the detector’s output remains below a specified threshold.
\begin{definition}
\label{def_alpha_stealth}
For a given time horizon $T$, an attack $\mathbf{a}$ of~\eqref{eq_system_bf_x} is $\alpha$-stealthy in $k=1,\cdots,T$ if
the following holds:
\begin{align}
    \exists\alpha\in\mathbb{R}_{+}\mathrm{~\st{}~}\|\mathbf{y}-\mathbf{y_{n}}\|^2 \leq \alpha    
\end{align}
where $\mathbf{y}$ is given by~\eqref{eq_system_bf_y} and $\mathbf{y_{n}}$ is by $\mathbf{y}$ of~\eqref{eq_system_bf_y} with $\mathbf{a}=0$.
\end{definition}

We summarize the key challenges in designing a general defense strategy against $\alpha$-stealthy network adversaries as follows:

\begin{enumerate}[label={\color{uiucblue}{(\alph*)}}] \item How can we adapt the regret-optimal control approach~\eqref{eq_regret_policy} to handle the $\alpha$-stealthy attacks defined in Definition~\ref{def_alpha_stealth}, instead of using $\|\mathbf{a}\| \leq 1$ (see Theorem~\ref{thm_regret_metric})?\label{item_problem1}
\item How can we minimize the metric from~\ref{item_problem1} using system-level parameterization for scalable, distributed control in large networks (see Proposition~\ref{proposition_sls_rstealthy_regret})?\label{item_problem2} \item How can we solve~\ref{item_problem2} for the network topology $\bm{\nu}$ in~\eqref{eq_output_feedback} in a computationally efficient way to optimally detect and mitigate $\alpha$-stealthy attacks (see Theorem~\ref{thm_convex_rank_constraint})?\label{item_problem3} \end{enumerate}
\section{Regret-Optimal Defense Against Stealthy Attacks}
\label{sec_regret}
Building on~\ref{item_problem1}, we formulate the following optimization problem to evaluate the worst-case regret under $\alpha$-stealthy attacks, as defined in Definition~\ref{def_alpha_stealth}:
\begin{align}
    \label{eq_regret_optimal}
    \sigma(\mathbf{u}) = \max_{\mathbf{a}}\regret_{J}(\mathbf{u},\mathbf{a})~\mathrm{\st{}}~\text{$\mathbf{a}$ is $\alpha$ stealthy}.
\end{align}
This formulation enables the simultaneous consideration of both regret-based robustness and the stealth characteristics of the attack sequence $a_k$ from\eqref{eq_continuous_initialx}, analogous to the output-to-output gain approach used in $\mathcal{H}_{\infty}$ robustness analysis~\cite{outputL2gain}.
\subsection{Optimal Non-Causal Control}
We define the performance objective $J(\mathbf{u},\mathbf{a})$ from~\eqref{eq_Jdefinition} as $J(\mathbf{u},\mathbf{a}) = \|\mathbf{z}\|^2+\|\mathbf{u}\|^2$, where $\mathbf{z}$ is the regulated output for the control performance given in~\eqref{eq_system_bf_z}, and $\mathbf{u}$ is the control input . The non-causal optimal performance $\mathcal{J}^*(\mathbf{a})$ as in $J^*(\mathbf{a})$ of~\eqref{eq_Jdefinition} (where $\mathcal{J}$ is introduced to distinguish it from the performance objective $J$ in Sec.~\ref{sec_pre_regret}) can be expressed as follows:
\begin{align}
    \label{eq_Jstar}
    \mathcal{J}^*(\mathbf{a}) = \min_{\mathbf{u}}\|\mathcal{E}\mathbf{u}+\mathcal{F}\mathcal{B}_a\mathbf{a}\|^2+ \|\mathbf{u}\|^2
\end{align}
where $\mathcal{E} = \mathcal{C}_z(\mathbb{I}-\mathcal{Z}\mathcal{A})^{-1}\mathcal{Z}\mathcal{B}_u+\mathcal{D}_{zu}$, $\mathcal{F} = \mathcal{C}_z(\mathbb{I}-\mathcal{Z}\mathcal{A})^{-1}$. This problem is clearly convex and can be solved as~\cite{regret_optimal_control,slsregret} 
\begin{align}
    \label{eq_Q_nc_def}
    \mathcal{J}^*(\mathbf{a}) =\mathbf{a}^{\top}\mathcal{B}_a^{\top} \mathcal{F}^{\top} (\mathcal{I}+\mathcal{E}\mathcal{E}^{\top})^{-1}\mathcal{F}\mathcal{B}_a \mathbf{a}=\mathbf{a}^{\top}\mathcal{Q}\mathbf{a}
\end{align}
where $\mathcal{Q}$ is defined appropriately. The optimal non-causal control policy $\mathbf{u}_{nc}$ of~\eqref{eq_Jdefinition} is then given as
\begin{align}
    \mathbf{u_{nc}}(\mathbf{a}) = - (\mathcal{I}+\mathcal{E}^{\top}\mathcal{E})^{-1}\mathcal{E}^{\top}\mathcal{F}\mathcal{B}_a\mathbf{a}.
\end{align}
\subsection{System Level Approach to Worst-Case Stealthy Attack}
Let $\mathbf{\Omega} = (\mathbf{R},\mathbf{N},\mathbf{M},\mathbf{L})$ for the system level parameterization~\eqref{eq_system_with_feedback}. Then the constraint of~\eqref{eq_regret_optimal} can be written as
\begin{align}
    \label{eq_stealth_con}    &\mathbf{a}^{\top}\Phi(\mathbf{\Omega})^{\top}\Phi(\mathbf{\Omega})\mathbf{a} \leq \alpha,\\
    &\Phi(\mathbf{\Omega})
    =
    \begin{bmatrix}
        \mathcal{C}_y(\mathbf{R}\mathcal{B}_a +\mathbf{N}\mathcal{D}_{ya})+\mathcal{D}_{ya}
    \end{bmatrix}= \begin{bmatrix}
        \mathcal{C}_y\mathbf{\phi}_x+\mathcal{D}_{ya}
    \end{bmatrix}\\&=
   \begin{bmatrix}       \mathcal{C}_y&0\end{bmatrix}\begin{bmatrix}\mathbf{\phi}_x\\\mathbf{\phi}_u\end{bmatrix} +\mathcal{D}_{ya}
\end{align}
by Definition~\ref{def_alpha_stealth}. Also, $J(\mathbf{u},\mathbf{a}) = \mathcal{J}(\mathbf{\Omega},\mathbf{a}) = \|\mathbf{z}\|^2+\|\mathbf{u}\|^2$  is given as
\begin{align}
    \label{eq_stealth_obj}
    \mathcal{J}(\mathbf{\Omega},\mathbf{a}) &= 
    \mathbf{a}^{\top}
    (\Psi(\mathbf{\Omega})^{\top}
    \Psi(\mathbf{\Omega})+ \mathbf{\Phi}_u(\mathbf{\Omega})^{\top}\mathbf{\Phi}_u(\mathbf{\Omega}))
    \mathbf{a} \\
    \Psi(\mathbf{\Omega})
    &=
    \begin{bmatrix}
        \mathcal{C}_z\mathbf{R}\mathcal{B}_a+\mathcal{D}_{zu}\mathbf{M}\mathcal{B}_a +\mathcal{C}_z\mathbf{N}\mathcal{D}_{ya}+\mathcal{D}_{zu}\mathbf{L}\mathcal{D}_{ya}
    \end{bmatrix}\\
    &= \begin{bmatrix}    \mathcal{C}_z\mathbf{\phi}_x+\mathcal{D}_{zu}\mathbf{\phi}_u
    \end{bmatrix}= \begin{bmatrix}    \mathcal{C}_z&\mathcal{D}_{zu}
    \end{bmatrix}\begin{bmatrix} \mathbf{\phi}_x\\\mathbf{\phi}_u\end{bmatrix}
\end{align}
and $\mathbf{\Phi}_u(\mathbf{\Omega})$ is as in \eqref{eq_system_with_feedback1}.
The metric in~\eqref{eq_regret_optimal} can now be formulated as the following quadratically constrained quadratic program (QCQP), using the system-level parameterization from~\eqref{eq_system_with_feedback} and~\eqref{eq_sls_conditions}:
\begin{align}
    \label{eq_stealth_qcqp}
    \varsigma(\mathbf{\Omega}) = \max_{\mathbf{a}}\mathbf{a}^{\top}(
    \Psi^{\top}
    \Psi+\Phi_u^{\top}\Phi_u-\mathcal{Q})
    \mathbf{a}~\mathrm{\st{}}~\mathbf{a}^{\top}\Phi^{\top}\Phi\mathbf{a} \leq \alpha.
\end{align}
For simplicity, the argument $\mathbf{\Omega}$ for $\Phi$, $\Psi$, and $\Phi_u$ is omitted here. The terms $\mathcal{Q}$, $\Phi$, $\Psi$, and $\Phi_u$ are defined in~\eqref{eq_Q_nc_def}, \eqref{eq_stealth_con}, \eqref{eq_stealth_obj}, and~\eqref{eq_system_with_feedback1}, respectively. QCQPs exhibit strong duality when Slater’s condition holds~\cite[pp. 653-658]{citeulike:163662},\cite{slsregret}, leading to the following theorem.
\begin{theorem}
\label{thm_regret_metric}
The QCQP for the regret-optimal metric for $\alpha$-stealthy attacks~\eqref{eq_stealth_qcqp} is equivalent to the following convex optimization problem:
\begin{align}
    \label{eq_regret_metric_convex}
    &\mu(\mathbf{\Omega}) = \min_{\lambda\geq 0}\lambda\alpha~\\
    &\mathrm{\st{}}~\lambda\Phi(\mathbf{\Omega})^{\top}\Phi(\mathbf{\Omega})-\Psi(\mathbf{\Omega})^{\top}\Psi(\mathbf{\Omega})-\Phi_u(\mathbf{\Omega})^{\top}\Phi_u(\mathbf{\Omega})+\mathcal{Q} \succeq 0~~~
\end{align}
and thus $\mu(\mathbf{\Omega}) = \varsigma(\Omega)$ for $\varsigma$ of~\eqref{eq_stealth_qcqp}.
\end{theorem}
\begin{proof}
The Lagrangian of the QCQP~\eqref{eq_stealth_qcqp} is given as follows:
\begin{align}
    L(\mathbf{a},\lambda) = \lambda(\mathbf{a}^{\top}\Phi^{\top}\Phi\mathbf{a}-\alpha)+\mathbf{a}^{\top}(-\Psi^{\top}\Psi-\Phi_u^{\top}\Phi_u+\mathcal{Q})\mathbf{a}
\end{align}
which gives the Lagrange dual function $g(\lambda)$ defined as
\begin{align}
    g(\lambda) = \min_{\mathbf{a}}L
    =
    \begin{cases}
    -\lambda\alpha & \text{$\lambda\Phi^{\top}\Phi-\Psi^{\top}\Psi-\Phi_u^{\top}\Phi_u+\mathcal{Q} \succeq 0$}\\
    -\infty & \text{otherwise}
    \end{cases}.
\end{align}
As a result, we get the following convex optimization problem:
\begin{align}
    \min_{\lambda\geq 0}\lambda\alpha~\mathrm{\st{}}~\lambda\Phi^{\top}\Phi-\Psi^{\top}\Psi-\Phi_u^{\top}\Phi_u+\mathcal{Q} \succeq 0.
\end{align}
Since the original QCQP~\eqref{eq_stealth_qcqp} is strictly feasible for $\mathbf{a} = 0$ when $\alpha\in\mathbb{R}_{+}$, strong duality holds~\cite[pp. 653-658]{citeulike:163662}.
\end{proof}

The metric introduced in~\eqref{eq_regret_metric_convex} provides a novel framework for quantifying worst-case regret in controllers, leveraging the system-level parameterization outlined in Sec.\ref{sec_sls}. Similar to the output-to-output gain discussed in\cite{outputL2gain}, this metric can be interpreted as a system gain from the $\alpha$-stealthy attack $\mathbf{a}$ in~\eqref{eq_system_bf_x} to the regret of the regulated performance output $\mathbf{z}$ in~\eqref{eq_system_bf_z}, thereby addressing problem~\ref{item_problem1}. Theorem~\ref{thm_regret_metric} introduces a more refined and flexible method for assessing system resilience against stealthy attacks through convex optimization, employing regret as an alternative to traditional system gains such as $\mathcal{H}_{\infty}$ and $\mathcal{H}_2$.

\subsection{Regret-Optimal Defense via Topology Switching}
Using the system-level parameterization in Sec.\ref{sec_sls}, we now design a feedback control policy to minimize the metric $\mu$ from\eqref{eq_regret_metric_convex} for regret-optimal stealthy attack mitigation.
\begin{proposition}
\label{proposition_sls_rstealthy_regret}
Consider the following optimization problem defined by the system-level parameterization $\mathbf{\Omega} = (\mathbf{R},\mathbf{M},\mathbf{N},\mathbf{L})$ of~\eqref{eq_system_with_feedback}:
 \begin{subequations}
\label{eq_prop_nonconvex}
\begin{align}
    &\mu^* = \min_{\mathbf{\Omega},\lambda}\lambda\alpha\label{eq_obj_prop}\textcolor{white}{\eqref{eq_obj_prop}} \\
    &\mathrm{\st{}}~\lambda\Phi(\mathbf{\Omega})^{\top}\Phi(\mathbf{\Omega})-\Psi(\mathbf{\Omega})^{\top}\!\Psi(\mathbf{\Omega})-\Phi_u(\mathbf{\Omega})^{\top}\!\Phi_u(\mathbf{\Omega})+\mathcal{Q} \succeq 0\label{eq_con_prop}\textcolor{white}{\eqref{eq_con_prop}} \\
    &\textcolor{white}{\mathrm{\st{}}~}
    \begin{bmatrix}
        \mathbb{I}-\mathcal{Z}\mathcal{A} & -\mathcal{Z}\mathcal{B}_u
    \end{bmatrix} 
    \begin{bmatrix}
        \mathbf{R} & \mathbf{N} \\
        \mathbf{M} & \mathbf{L}
    \end{bmatrix}
    =
    \begin{bmatrix}
        \mathbb{I} & \mathbb{O}
    \end{bmatrix} \label{eq_sls1_prop}\textcolor{white}{\eqref{eq_sls1_prop}} \\
    &\textcolor{white}{\mathrm{\st{}}~}
    \begin{bmatrix}
        \mathbf{R} & \mathbf{N} \\
        \mathbf{M} & \mathbf{L}
    \end{bmatrix}
    \begin{bmatrix}
        \mathbb{I}-\mathcal{Z}\mathcal{A} \\
        -\mathcal{C}_y
    \end{bmatrix}
    =
    \begin{bmatrix}
        \mathbb{I} \\
        \mathbb{O}
    \end{bmatrix} \label{eq_sls2_prop}\textcolor{white}{\eqref{eq_sls2_prop}}\\
    &\textcolor{white}{\mathrm{\st{}}~}\mathbf{\Omega} \in \mathfrak{S} (= \mathrm{set~for~system~level~constraints)},~\lambda\geq0\label{eq_sls3_prop}\textcolor{white}{\eqref{eq_sls3_prop}}
\end{align}
\end{subequations}
where $\mathcal{Q}$, $\Phi$, and $\Psi$ are as given in~\eqref{eq_Q_nc_def}, \eqref{eq_stealth_con}, and~\eqref{eq_stealth_obj}, respectively, and and other terms are defined in~\eqref{eq_system_bf}.
The output feedback control of~\eqref{eq_output_feedback} with $\mathbf{K} = \mathbf{K}^* = \mathbf
{L}^*-\mathbf{M}^*{\mathbf{R}^*}^{-1}\mathbf{N}^*$ achieves the desired response~\eqref{eq_system_with_feedback} and ensures regret-optimal resilience against $\alpha$-stealthy attacks (Definition~\ref{def_alpha_stealth}). The optimal solution $\mathbf{\Omega}^* = (\mathbf{R}^*,\mathbf{M}^*,\mathbf{N}^*,\mathbf{L}^*)$ guarantees regret-optimal performance. 

In particular, for the dynamic regret $\regret_{\mathcal{J}}(\mathbf{\Omega},\mathbf{a}) = \mathcal{J}(\mathbf{\Omega},\mathbf{a})-\mathcal{J}^*(\mathbf{a})$ where $\mathcal{J}^*$ and $\mathcal{J}$ are defined in~\eqref{eq_Jstar} and~\eqref{eq_stealth_obj}, the optimal value $\mu^*$ of~\eqref{eq_prop_nonconvex} satisfies \begin{align}
    \label{eq_prop_regret1}
     \regret_{\mathcal{J}}(\mathbf{\Omega}^*,\mathbf{a}) \leq \mu^*\mathrm{,~\forall\alpha\text{-}stealthy~attacks~\mathbf{a}}
\end{align}
Moreover, for the worst-case $\alpha$-stealthy attack $\mathbf{d_{wc}}(\mathbf{\Omega})$, which minimizes~\eqref{eq_stealth_qcqp} (the optimal value given in Theorem~\ref{thm_regret_metric}), we have
\begin{align}
    \label{eq_prop_regret2}
     \mu^* = \regret_{\mathcal{J}}(\mathbf{\Omega}^*,\mathbf{a}^*) \leq \regret_{\mathcal{J}}(\mathbf{\Omega},\bar{\mathbf{a}})\mathrm{,~\forall\mathbf{\Omega}~with~\eqref{eq_sls_conditions}}~~~
\end{align} 
where $\mathbf{a}^* = \mathbf{d_{wc}}(\mathbf{\Omega}^*)$ and $\bar{\mathbf{a}} = \mathbf{d_{wc}}(\mathbf{\Omega})$.
\end{proposition}

\begin{proof}
As discussed in Lemma~\ref{lemma_sls}, the conditions~\eqref{eq_sls1_prop} and~\eqref{eq_sls2_prop} ensure that the feedback controller with $\mathbf{K} = \mathbf{K}^* = \mathbf
{L}^*-\mathbf{M}^*{\mathbf{R}^*}^{-1}\mathbf{N}^*$ achieves the desired response~\eqref{eq_system_with_feedback}. Since the problem~\eqref{eq_prop_nonconvex} is formulated to find $\mathbf{\Omega}$ that minimizes the metric $\mu$ of~\eqref{eq_regret_metric_convex} in Theorem~\ref{thm_regret_metric}, the relations~\eqref{eq_prop_regret1} and~\eqref{eq_prop_regret2} naturally follow by construction of $\varsigma$ in~\eqref{eq_stealth_qcqp}. 
\end{proof}

Due to the affine nature of the system-level parameterization in Lemma~\ref{lemma_sls}, the system-level constraints $\mathfrak{S}$ can accommodate a wide range of system properties, including temporal and spatial locality, distributed control, and scalability~\cite{sls_ARC,sls_output_feedback_control,sls_tutorial}. In particular, for networked control systems, the control gain can be adjusted by switching the topology as in~\eqref{eq_output_feedback}, and distributed control constraints can be incorporated directly into~\eqref{eq_system_with_feedback} in affine terms based on the network topology. This leads to the following assumption.
\begin{assumption}
\label{assump_affine_sls_topology}
Given the history of topologies of $\bm{\nu}$~\eqref{eq_system_with_feedback} and their adjacency matrices, we can define a convex constraint $\mathbf{\Omega} \in \mathfrak{C}(\bm{\nu})$ for restricting $\mathbf{\Omega}$ to be in a set consistent with the network topologies and the system's causality.
\end{assumption}

However, despite using $\mathfrak{S} = \mathfrak{C}(\bm{\nu})$ from Assumption~\ref{assump_affine_sls_topology}, the optimization problem~\eqref{eq_prop_nonconvex} in Proposition~\ref{proposition_sls_rstealthy_regret} remains nonlinear. Before proceeding, we introduce the following lemma useful for convexification (see also, \eg{},~\cite{BoydRankRelaxation,rank_constrained_kyotoU}).
\begin{lemma}
\label{lemma_rank}
For any real column vector $a\in\mathbb{R}^b$ and real symmetric matrix $S\in\mathbb{S}^{b\times b}$, we have the following:
\begin{align}
    \label{eq_rank_equivalence}
    \{(a,S)|S=aa^{\top}\} = \left\{(a,S)\left|\begin{bmatrix}
        S & a \\
        a^{\top} & 1
    \end{bmatrix} \succeq 0~\&~\rank(S)=1\right.\right\}~~~
\end{align}
where $\rank(S)$ denotes the matrix rank of $S$.
\end{lemma}
\begin{proof}
Let $\mathrm{(LHS)}=\mathfrak{L}$ and $\mathrm{(RHS)}=\mathfrak{R}$. 

($\mathfrak{L}\subset\mathfrak{R}$): Since $\rank(S) = \rank(aa^{\top}) = 1$ and $S=aa^{\top}\Rightarrow S\succeq aa^{\top}$, applying Schur's complement lemma~\cite[pp. 650-651]{citeulike:163662} shows $\mathfrak{L}\subset\mathfrak{R}$.

($\mathfrak{R}\subset\mathfrak{L}$): Since $S$ is real symmetric, we can always write $S$ as $S = V\diag(\varpi_1,\cdots,\varpi_b)V^{\top}$, where $V$ is an orthogonal matrix of the orthonormal eigenvectors and $\varpi_i$ are the real eignenvalues of $S$. Also, since $\rank(S) = 1$, $S$ has at most one non-zero eigenvalue $\varpi$, which implies $S = \varpi vv^{\top}$ for the corresponding unit eigenvector $v$. Defining $a$ as $a = \sqrt{\varpi} v$ shows $\mathfrak{R}\subset\mathfrak{L}$.
\end{proof}

The following theorem provides one way to reformulate the problem~\eqref{eq_prop_nonconvex} of Proposition~\ref{proposition_sls_rstealthy_regret} into an almost equivalent optimization problem in a computationally efficient form. 
\begin{theorem}
\label{thm_convex_rank_constraint}
Let Assumption~\ref{assump_affine_sls_topology} hold and let $\mathfrak{S} = \mathfrak{C}(\bm{\nu})$ as in~\eqref{eq_prop_nonconvex}. Consider the following optimization problem, where the objective and constraints are convex, except for the rank constraints in~\eqref{eq_con_thmconvexrank4}: 
\begin{subequations}
\label{eq_thm_convexrank}
\begin{align}
    &\bar{\mu}^* = \min_{\mathbf{\Omega},\bm{\lambda},\bm{\mathcal{X}},\mathbf{\Lambda}}\lambda\alpha\label{eq_thm_convexrank_obj}\textcolor{white}{\eqref{eq_thm_convexrank_obj}} \\
    &\mathrm{\st{}}~
    \begin{bmatrix}
        \lambda_{\mathrm{inv}}\mathcal{Q}+\ell(\bm{\mathcal{X}}) & \Psi(\mathbf{\Omega})^{\top}&\Phi_u(\mathbf{\Omega})^{\top} \\
        \Psi(\mathbf{\Omega}) & \lambda\mathbb{I} &\mathbb{O}\\
        \Phi_u(\mathbf{\Omega}) & \mathbb{O}&\lambda\mathbb{I} 
    \end{bmatrix} \succeq 0,~\eqref{eq_sls1_prop},~\eqref{eq_sls2_prop} \label{eq_con_thmconvexrank1}\textcolor{white}{\eqref{eq_con_thmconvexrank1}} \\
    &\textcolor{white}{\mathrm{\st{}}~}
    \begin{bmatrix}
        \bm{\mathcal{X}} & \matvec(\Phi(\mathbf{\Omega})) \\
        \matvec(\Phi(\mathbf{\Omega}))^{\top} & 1
    \end{bmatrix} \succeq 0,~
    \begin{bmatrix}
        \mathbf{\Lambda} & \bm{\lambda} \\
        \bm{\lambda}^{\top} & 1
    \end{bmatrix} \succeq 0, \label{eq_con_thmconvexrank2}\textcolor{white}{\eqref{eq_con_thmconvexrank2}}\\
    &\textcolor{white}{\mathrm{\st{}}~}\mathbf{\Omega} \in \mathfrak{C}(\bm{\nu}),~\bm{\lambda} = [\lambda,\lambda_{\mathrm{inv}}]^{\top},~\lambda > 0,~\mathbf{\Lambda}_{12} = \mathbf{\Lambda}_{21} = 1 \label{eq_con_thmconvexrank3}\textcolor{white}{\eqref{eq_con_thmconvexrank3}} \\
    &\textcolor{white}{\mathrm{\st{}}~}\rank(\bm{\mathcal{X}}) = 1, \rank(\mathbf{\Lambda}) = 1 \label{eq_con_thmconvexrank4}\textcolor{white}{\eqref{eq_con_thmconvexrank4}}
\end{align}
\end{subequations}
where $\ell$ is a linear function of the decision variable $\bm{\mathcal{X}}$ defined by $\ell(\bm{\mathcal{X}}) = \Phi(\mathbf{\Omega})^{\top}\Phi(\mathbf{\Omega})$, and $\matvec(\Phi(\mathbf{\Omega}))$ denotes the vectorization of the matrix $\Phi(\mathbf{\Omega})$. Other notations follow those in Proposition~\ref{proposition_sls_rstealthy_regret}. If $\lambda^* \neq 0$ for the optimizer $(\lambda^*,\Omega^*)$ of~\eqref{eq_prop_nonconvex}, which is the case when the optimal regret $\mu^* = \regret_{\mathcal{J}}(\mathbf{\Omega}^*,\mathbf{a}^*) > 0$ in~\eqref{eq_prop_regret2}, then we have
\begin{align}
    \label{eq_mu_equal}
    \bar{\mu}^* = \mu^*.
\end{align}
If $\lambda^* = 0$, which is the case when the optimal regret $\mu^* = \regret_{\mathcal{J}}(\mathbf{\Omega}^*,\mathbf{a}^*) = 0$, then we have
\begin{align}
    \label{eq_mu_not_equal}
    \bar{\mu}^* \geq \mu^*.
\end{align}
\end{theorem}
\begin{proof}
If $\lambda^* \neq 0$ in~\eqref{eq_prop_nonconvex}, \ie{}, $\mu^* = \regret_{\mathcal{J}}(\mathbf{\Omega}^*,\mathbf{a}^*) > 0$ in~\eqref{eq_prop_regret2}, the constraint~\eqref{eq_con_prop} can be equivalently expressed as
\begin{align}
    &\begin{bmatrix}        \lambda_{\mathrm{inv}}\mathcal{Q}+\mathcal{P} & \Psi(\mathbf{\Omega})^{\top}&\Phi_u(\mathbf{\Omega})^{\top} \\
        \Psi(\mathbf{\Omega}) & \lambda\mathbb{I} &\mathbb{O}\\
        \Phi_u(\mathbf{\Omega}) & \mathbb{O}&\lambda\mathbb{I} 
    \end{bmatrix} \label{eq_nonlinaer_lmi}
    \succeq 0, \\
    &\lambda \lambda_{\mathrm{inv}} = 1,~\mathcal{P} = \Phi(\mathbf{\Omega})^{\top}\Phi(\mathbf{\Omega}) \label{eq_nonlinaer_quadratic}
\end{align}
using Schur's complement lemma~\cite[pp. 650-651]{citeulike:163662}, which is nonlinear due to the equality constraints~\eqref{eq_nonlinaer_quadratic}. Still, since the nonlinearity is at most quadratic/bilinear, these constraints can be written as a function of additional decision variables (\ie{}, liftings) for $\mathbf{\Lambda} = \bm{\lambda}\bm{\lambda}^{\top}$ and $\bm{\mathcal{X}} = \matvec(\Phi(\mathbf{\Omega}))\matvec(\Phi(\mathbf{\Omega}))^{\top}$, where $\bm{\lambda} = [\lambda,\lambda_{\mathrm{inv}}]^{\top}$. This observation leads to the following equivalent constraints of~\eqref{eq_nonlinaer_quadratic} due to~\eqref{eq_rank_equivalence} of Lemma~\ref{lemma_rank}:
\begin{align}
    \eqref{eq_nonlinaer_quadratic} \Leftrightarrow& 
    \begin{bmatrix}
        \bm{\mathcal{X}} & \matvec(\Phi(\mathbf{\Omega})) \\
        \matvec(\Phi(\mathbf{\Omega}))^{\top} & 1
    \end{bmatrix} \succeq 0,~
    \begin{bmatrix}
        \mathbf{\Lambda} & \bm{\lambda} \\
        \bm{\lambda}^{\top} & 1
    \end{bmatrix} \succeq 0, \\
    \textcolor{white}{\eqref{eq_nonlinaer_quadratic} \Leftrightarrow}&\rank(\bm{\mathcal{X}}) = 1,~\rank(\mathbf{\Lambda}) = 1.
\end{align}
Rewriting the constraints~\eqref{eq_nonlinaer_lmi} and~\eqref{eq_nonlinaer_quadratic} using these additional lifted variables completes the proof for~\eqref{eq_mu_equal}.

The relation~\eqref{eq_mu_not_equal} can be immediately obtained by the fact that $\{\lambda\in\mathbb{R}|\lambda > 0\} \subset \{\lambda\in\mathbb{R}|\lambda \geq 0\}$.
\end{proof}

Since the optimization problem derived in Theorem~\ref{thm_convex_rank_constraint} is convex except for the rank constraint,  we can apply various convex rank minimization techniques from the literature, including, but not limited to,~\cite{BoydRankRelaxation,rank_constrained_kyotoU,rank_constrained_boyd,rank_constrained_microsoft,rank_constrained_pourdue_ran}.

\subsection{Relaxed Regret-Optimal Control}
As in the relaxed, suboptimal $\mathcal{H}_{\infty}$ control (see, e.g.,\cite{256331}), we can consider a suboptimal, bounded regret-optimal gain for stealthy attacks in Proposition~\ref{proposition_sls_rstealthy_regret}, which results in a simpler optimization problem as follows.
\begin{corollary}
\label{corollary_suboptimal_hinfinity}
If we consider a relaxed problem with a fixed $\lambda = \bar{\lambda}\in\mathbb{R}_{+}$ in~\eqref{eq_prop_nonconvex} of Proposition~\ref{proposition_sls_rstealthy_regret}, then the problem~\eqref{eq_thm_convexrank} simplifies to the following feasibility problem:
\begin{subequations}
\label{eq_corollary_convexrank}
\begin{align}
    &\mathrm{Find}~(\mathbf{\Omega},\bm{\mathcal{X}})\label{eq_corollary_convexrank_obj}\textcolor{white}{\eqref{eq_corollary_convexrank_obj}} \\
    &\mathrm{\st{}}~
    \begin{bmatrix}        \mathcal{Q}+\lambda\ell(\bm{\mathcal{X}}) & \Psi(\mathbf{\Omega})^{\top}&\Phi_u(\mathbf{\Omega})^{\top} \\
        \Psi(\mathbf{\Omega}) & \mathbb{I} &\mathbb{O}\\
        \Phi_u(\mathbf{\Omega}) & \mathbb{O}&\mathbb{I} 
    \end{bmatrix} \succeq 0,~\eqref{eq_sls1_prop},~\eqref{eq_sls2_prop} \label{eq_corollary_convexrank1}\textcolor{white}{\eqref{eq_corollary_convexrank1}} \\
    &\textcolor{white}{\mathrm{\st{}}~}
    \begin{bmatrix}
        \bm{\mathcal{X}} & \matvec(\Phi(\mathbf{\Omega})) \\
        \matvec(\Phi(\mathbf{\Omega}))^{\top} & 1
    \end{bmatrix} \succeq 0,~\mathbf{\Omega} \in \mathfrak{C}(\bm{\nu}),\label{eq_corollary_convexrank2}\textcolor{white}{\eqref{eq_corollary_convexrank2}} \\
    &\textcolor{white}{\mathrm{\st{}}~}\rank(\bm{\mathcal{X}}) = 1 \label{eq_corollary_convexrank3}\textcolor{white}{\eqref{eq_corollary_convexrank3}}
\end{align}
which is convex except for the rank condition~\eqref{eq_corollary_convexrank3}.
\end{subequations}
\end{corollary}
\begin{proof}
This follows from Theorem~\ref{thm_convex_rank_constraint} by fixing $\lambda=\bar{\lambda}$.
\end{proof}

Also, dropping the rank constraints in the problems~\eqref{eq_thm_convexrank} and~\eqref{eq_corollary_convexrank} results in a convex optimization problem with a well-known relaxation approach.
\begin{corollary}\label{corollary_schor}
Applying Shor's relaxation~\cite{shor_original} to the problem~\eqref{eq_prop_nonconvex} of Proposition~\ref{proposition_sls_rstealthy_regret} yields the problem~\eqref{eq_thm_convexrank} of Theorem~\ref{thm_convex_rank_constraint} without the rank constraints~\eqref{eq_con_thmconvexrank4}. The solution of this problem $\bar{\mu}^*_{\mathrm{shor}}$ gives the lower bound of $\mu^*$ in~\eqref{eq_obj_prop}, \ie{},
\begin{align}
    \bar{\mu}^*_{\mathrm{shor}} \leq \mu^*.
\end{align}
\end{corollary}
\begin{proof}
This follows from~\eqref{eq_rank_equivalence} of Lemma~\ref{lemma_rank}, which indicates that dropping the rank constraint there results in Shor's relaxation. See also, \eg{}, \cite[pp. 220-225]{Shor1998Book},\cite{BoydRankRelaxation}.
\end{proof}
\section{Numerical example}
\label{sec_simulation}
This section presents the results of numerical simulation to demonstrate the efficacy of our proposed method. 

Two approaches are used: 1) Fix $\lambda$ and solve the feasibility problem in Corollary~\ref{corollary_suboptimal_hinfinity} using iterative rank minimization~\cite{rank_constrained_kyotoU}, adjusting $\lambda$ via linear search until a satisfactory sub-optimal solution is found. 2) Fix $\lambda$ and solve the relaxed problem in Corollary~\ref{corollary_schor}, compute the worst-case regret cost \eqref{eq_regret_metric_convex} under stealthy attacks, and iterate with a linear search on $\lambda$ until the cost \eqref{eq_regret_metric_convex} is sub-optimally minimized.

Using the above, we implement our proposed controller on a spring-damper system with two masses and a sampling time of $T_s = 0.5$ s, with the stealthiness measure set to $\alpha = 0.1$. We compare its performance to the classical $\mathcal{H}_\infty$ robust controller, with the results shown in this section. A comparison with the typical regret-optimal controller, consistent with~\eqref{eq_thm_convexrank2}, is omitted for brevity, as its performance under stealthy attacks closely resembles that of $\mathcal{H}_\infty$. Our formulation also allows the implementation of controllers resilient to stealthy attacks using any quadratic cost function, such as $\mathcal{H}_\infty$ cost function, along with stealthiness constraints, which would prioritize robustness over adaptability.

The classical $\mathcal{H}_\infty$ controller, which isn't designed specifically for stealthy attacks, solves the following problem:
\begin{subequations}
\label{eq_thm_convexrank2}
\begin{align}
    &\min_{\mathbf{\Omega}\in \mathfrak{C}(\bm{\nu})}\max_{\|\mathbf{a}\|\leq 1} \mathcal{J}(\mathbf{\Omega},\mathbf{a}) \\
    &= \min_{\mathbf{\Omega},\bm{\lambda}}\lambda\alpha\label{eq_thm_convexrank_obj2}\textcolor{white}{\eqref{eq_thm_convexrank_obj2}} \\
    &\mathrm{\st{}}~
    \begin{bmatrix}
        \lambda\mathcal{I} & \Psi(\mathbf{\Omega})^{\top}&\Phi_u(\mathbf{\Omega})^{\top} \\
        \Psi(\mathbf{\Omega}) & \mathbb{I} &\mathbb{O}\\
        \Phi_u(\mathbf{\Omega}) & \mathbb{O}&\mathbb{I} 
    \end{bmatrix} \succeq 0,~\eqref{eq_sls1_prop},~\eqref{eq_sls2_prop} \label{eq_con_thmconvexrank11}\textcolor{white}{\eqref{eq_con_thmconvexrank11}} \\
    &\textcolor{white}{\mathrm{\st{}}~}\mathbf{\Omega} \in \mathfrak{C}(\bm{\nu})\label{eq_con_thmconvexrank33}\textcolor{white}{\eqref{eq_con_thmconvexrank33}} 
\end{align}
\end{subequations}
where $\mathcal{J}(\mathbf{\Omega},\mathbf{a}) = 
    \mathbf{a}^{\top}
    (\Psi(\mathbf{\Omega})^{\top}
    \Psi(\mathbf{\Omega})+ \mathbf{\Phi}_u(\mathbf{\Omega})^{\top}\mathbf{\Phi}_u(\mathbf{\Omega}))
    \mathbf{a} $ as in \eqref{eq_stealth_obj}.
 
We compare the worst-case regret cost of $\mathcal{H}_\infty$ and our controller under stealthy attacks, as quantified by \eqref{eq_stealth_qcqp}, \eqref{eq_regret_metric_convex}. Table \ref{table:controller_costs} shows that our controller consistently outperforms $\mathcal{H}_\infty$ as the horizon increases.

\begin{figure}
    \centering
        \includegraphics[width=0.8\linewidth]{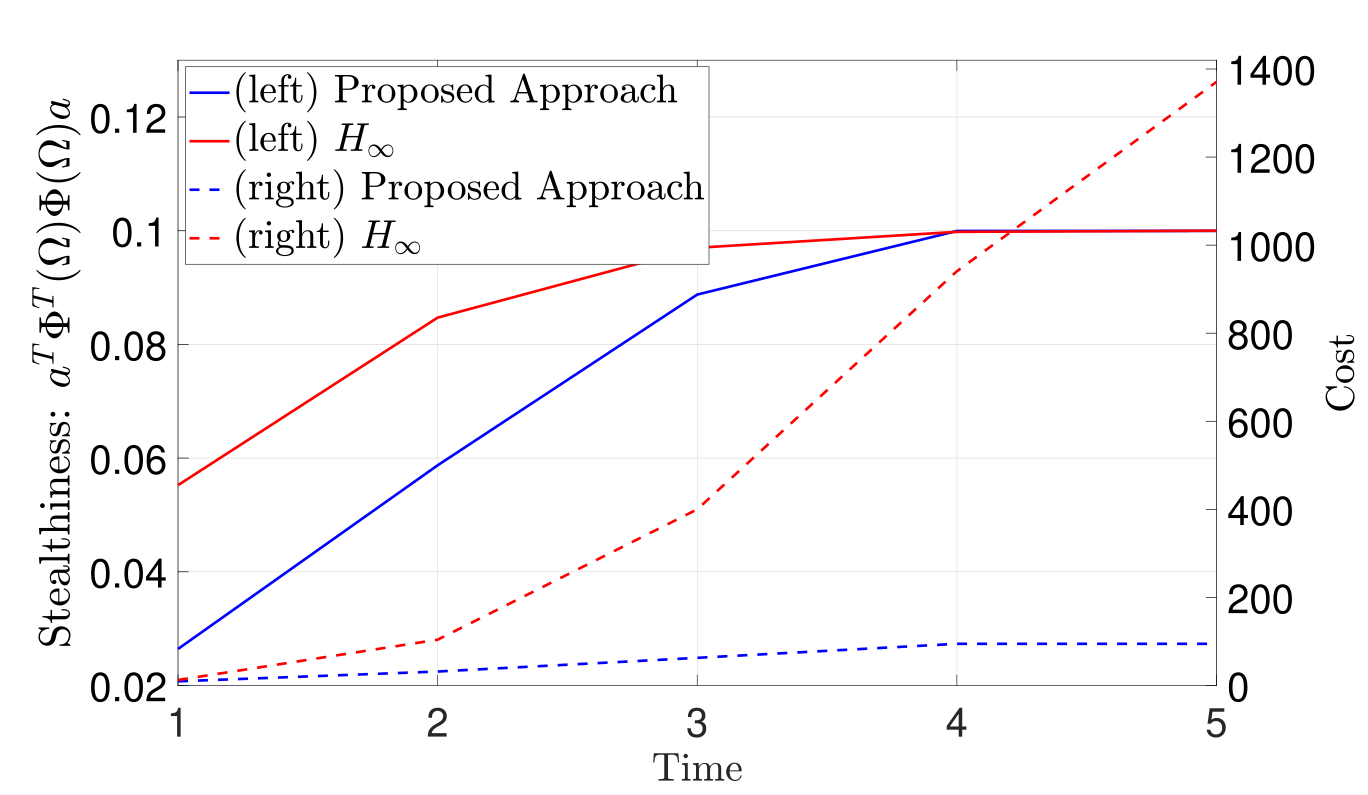}\vspace{-8pt}
        \caption{(Left) Stealthiness of the attacks measured as $\|y-y_n\|^2=a^\top \Phi(\Omega)^\top \Phi(\Omega) a$, for $\mathcal{H}_\infty$ and our controller. Both attacks satisfy the constraint: $\|y-y_n\|^2\leq 0.1$. (Right): Worst-case regret cost (\ref{eq_stealth_qcqp}). Our controller achieves a 15$\times$ lower cost compared to $\mathcal{H}_\infty$. }
        \label{fig:comparison_fig1}\vspace{-10pt}
\end{figure}

\begin{figure}
        \centering
        \includegraphics[width=0.8\linewidth]{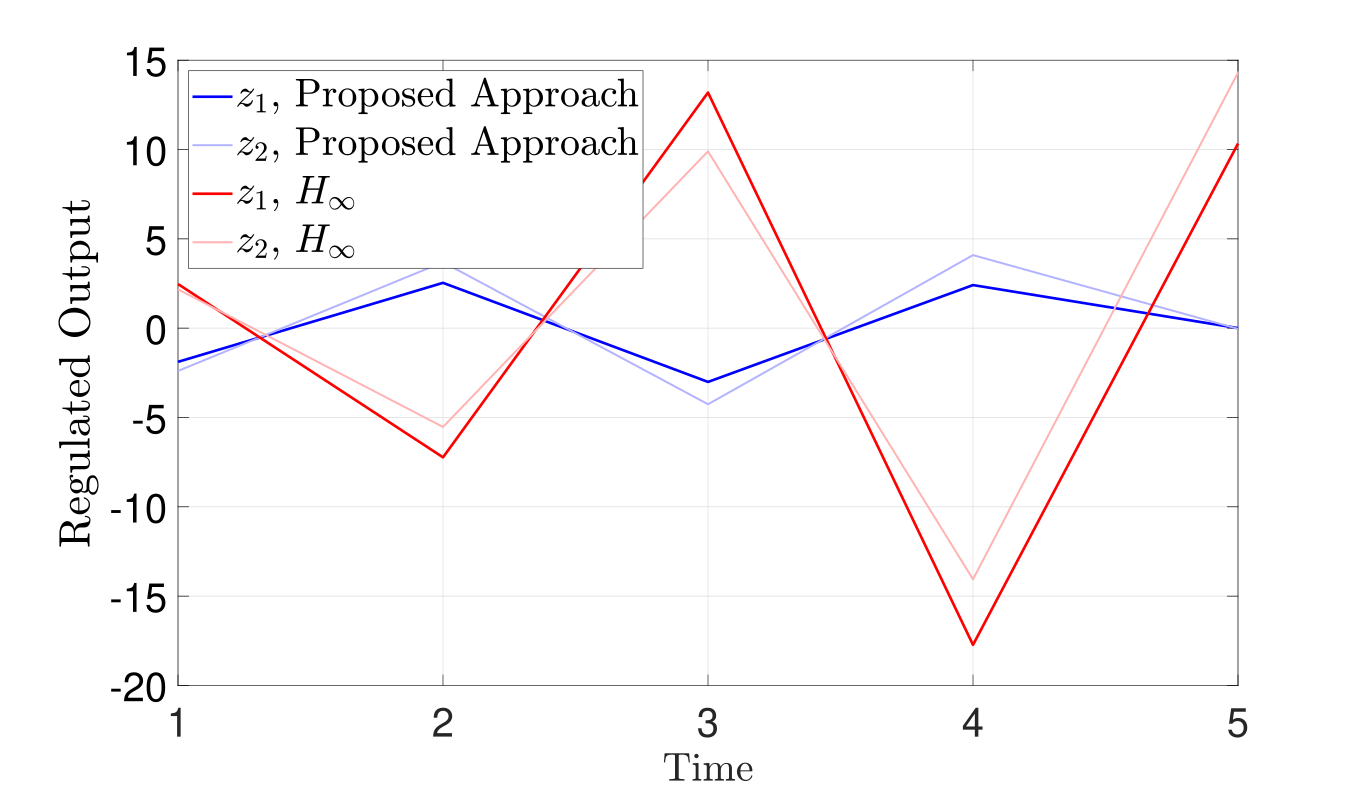}\vspace{-8pt}
        \caption{Regulated output $z$ (\ref{eq_system_bf_z}) under $\mathcal{H}_\infty$ and our controller. For a horizon of $T=5$, our controller manages to keep the regulated output near 0, while the regulated output under $\mathcal{H}_\infty$ shows big oscillations.}
        \label{fig:comparison_fig2}
    \vspace{-10pt}
    \label{fig:comparison}
\end{figure}

We also compute the worst-case stealthy attacks for each controller and apply them to our system \eqref{eq_system_bf}. Figure \ref{fig:comparison_fig1} shows that these attacks remain $\alpha$-stealthy ($\alpha=0.1$), with our controller achieving a 15x lower cost than $\mathcal{H}_\infty$. Figure \ref{fig:comparison_fig2} illustrates that only our controller regulates the output near zero. For computational efficiency, we chose a horizon of $T=5$. A model predictive control scheme, however, can be used to manage longer horizons in practice.

This example shows the potential of our system-level approach in detecting and mitigating stealthy attacks with regret-optimal strategies. Again, as will be detailed in the next section, the framework extends to a wider class of constrained controllers, including predictive control and online learning, while maintaining a convex formulation. 
\begin{table}[htbp] 
    \centering
    \caption{Worst-case regret under stealthy attacks as in (\ref{eq_stealth_qcqp}) for different time horizons (T=2, T=5).}\vspace{-8pt}
    \setlength{\tabcolsep}{2.5pt}
    \begin{tabular}{lcc}
        \toprule
        \textbf{} & \textbf{T=2} & \textbf{T=5} \\
        \midrule
        \texttt{$\mathcal{H}_\infty$ controller} & 41.38 & 1312.6 \\
        \texttt{Proposed approach} & 10.19 & 87.10\\
        \texttt{Improvement Factor: $\frac{\text{$\mathcal{H}_\infty$(Cost)}}{\text{Proposed approach (Cost)}}$} & \textbf{4.02} & \textbf{15.07}\\
        \bottomrule
    \end{tabular}
    \label{table:controller_costs}\vspace{-10pt}
\end{table}


\section{Potential Extensions}
\label{sec_extension}
This section explores various extensions of the topology switching strategy from Proposition~\ref{proposition_sls_rstealthy_regret} and Theorem~\ref{thm_convex_rank_constraint}, highlighting the broad applicability of our regret-optimal metric in more general problem settings under stealthy attacks. The purpose of this section is just to emphasize and outline the potential research directions that this paper would open up.

\subsection{Sparsity on System Responses}
If the convex constraint $\mathbf{\Omega} \in \mathfrak{C}(\bm{\nu})$ of~\eqref{eq_con_thmconvexrank3} in Theorem~\ref{thm_convex_rank_constraint} includes sparsity constraints due to, \eg{}, distributed communications of the network, we could use the following corollary for additional simplification.
\begin{corollary}
\label{eq_corollary_sparcity}
Suppose that we have sparsity constraints for $\mathbf{\Omega}$, which leads to linear constraints of the form $\matvec(\mathbf{\Phi}(\mathbf{\Omega}))_i = 0$. Then all the entries in the $i$th column and $i$th row of the matrix $\bm{\mathcal{X}}$ of~\eqref{eq_thm_convexrank} in Theorem~\ref{thm_convex_rank_constraint} are also equal to zero.
\end{corollary}
\begin{proof}
This follows from the definition of $\bm{\mathcal{X}}$ introduced in the proof of Theorem~\ref{thm_convex_rank_constraint}.
\end{proof}
Another potential direction for addressing sparsity is considering spatial regret (SpRegret) defined in \cite{martinelli2023closing} as the regret between the cost of a topology $\mathbf{\Omega}$ which belongs to a sparse set $\mathfrak{C}(\bm{\nu})$, and that of a topology $\hat{\mathbf{\Omega}}$ which belongs to a denser set $\mathfrak{C}(\bm{\hat{\nu}})$: SpRegret($\mathbf{\Omega}$,$\hat{\mathbf{\Omega}}$)= $\max_{\|a\| \leq 1} J(\mathbf{\Omega},a)-J(\hat{\mathbf{\Omega}},a)$. Under no further constraints, ideally, the constraint set should satisfy a quadratic invariance condition to ensure the optimality of the control strategy. In cases where it is not, there exist methods that allow for the synthesis of the controller even when the set is not Quadratic Invariant (QI). As in \cite{martinelli2023closing}, we could use the denser topology as the one that satisfies the QI condition.

We could further extend the notion of regret to spatiotemporal regret by defining it as the difference in costs of a sparse topology and of a denser topology that also knows the future attacks. In this case, the challenge we face is ensuring that the problem is well-posed ($\text{i.e., regret}>0$) and finding the closest QI, noncausal sparse set.
\subsection{Data-Driven Control}
The system level parameterization is also useful for the case where we have access only to the system's input and output data $u_{[0,T_\mathrm{data}-1]} = \{u_0,\cdots,u_{T_\mathrm{data}-1}\}$ and $y_{[0,T_\mathrm{data}-1]} = \{y_0,\cdots,y_{T_\mathrm{data}-1}\}$ of the system~\eqref{eq_system} with $a_k=0$. Let $\mathcal{H}(\sigma_{[0,j]})$ be the Hankel matrix of depth $D$, associated with the signal $\sigma_{[0,T_{\mathrm{data}}-1]}$ defined as follows:
\begin{align}
    \label{eq_Hankel}
    \mathcal{H}(\sigma_{[0,T_{\mathrm{data}}-1]}) = 
    \begin{bmatrix}
        \sigma_0 & \cdots & \sigma_{T_{\mathrm{data}}-D} \\
        \vdots & \ddots & \vdots \\
        \sigma_{D-1} & \cdots & \sigma_{T_{\mathrm{data}}-1}
    \end{bmatrix}
\end{align}
where $D \leq T_{\mathrm{data}}$.
\begin{definition}
\label{def_persistently_exciting}
The signal $\sigma_{[0,T_{\mathrm{data}}-1]}$ is persistently exciting of order $D$ if the Hankel matrix $\mathcal{H}(\sigma_{[0,T_{\mathrm{data}}-1]})$~\eqref{eq_Hankel} has full row rank.
\end{definition}

The Hankel matrix of~\eqref{eq_Hankel} leads to the following lemma essential in the data-driven system level synthesis~\cite{Willems_fundamental,Willems_fundamental_state_space}.
\begin{lemma}
\label{lemma_Willems}
Suppose that the input data $u_{[0,T_\mathrm{data}-1]}$ is persistently exciting of order $n+D$ as in Definition~\ref{def_persistently_exciting}, where $n$ is the number of the system state $x$ of~\eqref{eq_continuous_initialx}. Suppose also that the system~\eqref{eq_continuous_initialx} is time-invariant and controllable. Then $(\bar{u}_{[0,D-1]},\bar{y}_{[0,D-1]})$ is a valid input/output trajectory of~\eqref{eq_system} with $a_k = 0$ if and only if $\exists g\in\mathbb{R}^{T_\mathrm{data}-D+1}$ \st{}
\begin{align}
    \label{eq_Williems}
    \begin{bmatrix}
        u_{[0,T_\mathrm{data}-1]} \\
        y_{[0,T_\mathrm{data}-1]}
    \end{bmatrix}
    =
    \begin{bmatrix}
        \mathcal{H}(u_{[0,T_{\mathrm{data}}-1]}) \\
        \mathcal{H}(y_{[0,T_{\mathrm{data}}-1]})
    \end{bmatrix}g.
\end{align}
\end{lemma}
\begin{proof}
    See,~\eg{},~\cite{Willems_fundamental,Willems_fundamental_state_space}.
\end{proof}

It is shown in~\cite{data_driven_sls} that $g$ of~\eqref{eq_Williems} in Lemma~\ref{lemma_Willems} can be characterized by the system level parameterization of~\eqref{eq_system_with_feedback} for the case of full-state feedback. Its robust properties allow for extending this idea to perturbed trajectory data, as also shown in~\cite{data_driven_sls}. These observations imply the great potential of our approach to model-free settings as in~\cite{data_driven_sls,data_sls_mpc,sls_data_predictive}.
\subsection{Additional Constraints}
\subsubsection{Performance Constraints}
We can clearly see that the control input $\mathbf{u}$ in~\eqref{eq_system_with_feedback} of Sec.~\ref{sec_sls} is linear in the system level parameters $\mathbf{\Omega}$; the constraint
\begin{align}
    \|\mathbf{u}\| = \|\mathbf{M}\mathbf{d_x}+\mathbf{D}\mathbf{d_y}\|_* \leq u_{\max}
\end{align}
where $u_{\max}\in\mathbb{R}_{+}$ is some constant, is always a convex constraint for any norm $*$. We can also consider other performance requirements as convex constraints on top of the problems~\eqref{eq_thm_convexrank} of Theorem~\ref{thm_convex_rank_constraint} and~\eqref{eq_corollary_convexrank} of Corollary~\ref{eq_corollary_convexrank}, the examples of which are given in~\cite{sls_ARC,sls_tutorial,sls_output_feedback_control}.
\subsubsection{Localizability}
\label{sec_localizability}
For large-scale networks, we could apply spatiotemporal constraints to further reduce the computational burden thanks to the system level parameterization. In particular, if we assume that 1) the closed-loop disturbance responses have finite-impulse responses and that 2) the effects of disturbances are felt only in local neighborhoods of each agent, then we can decompose the problems~\eqref{eq_thm_convexrank} of Theorem~\ref{thm_convex_rank_constraint} into small subproblems that can be solved locally~\cite{sls_ARC,sls_tutorial,sls_output_feedback_control}. This allows the model predictive control-like formulation of the system level synthesis~\cite{slsmpc}.
\subsubsection{Controllability, Observability, and Sensitivity}
As discussed in~\cite{sls_output_feedback_control,localLQR,localLQG}, the localizability of Sec.~\ref{sec_localizability} allows for verifying the observability and controllability conditions using a set of affine constraints, which can be readily added to the problem~\eqref{eq_thm_convexrank} in Theorem~\ref{thm_convex_rank_constraint} while preserving its convex structure. The minimum attack sensitivity (or detectability in the original paper~\cite{attack_bullo}) could potentially be considered directly as a constraint in this problem using a similar approach. Alternatively, we could enforce the sensitivity constraint by using the $\mathcal{H}_{-}$ index~\cite{security_metric_book,Hsubobserver,LMIsensitivity,secure_control,outputL2gain} as in~\cite{topology_switching}, with the same lifting procedure of Theorem~\ref{thm_convex_rank_constraint}.
\subsubsection{Stability}
For infinite-horizon problems, we can still ensure the internal stability of our formulation in a convex manner as discussed in~\cite{sls_ARC,sls_tutorial,sls_output_feedback_control} and in~\cite{jing1} for adversarial settings. Even in our finite-horizon setting, we could still enforce the Lyapunov-type stability constraint as in~\cite{topology_switching} due to the affine nature of the system level parameterization as to be implied in the numerical example.

\section{Conclusion}
This paper introduced a system-level, regret-optimal framework for designing linear systems that are resilient to stealthy attacks and disturbances. We equivalently reformulated the nonlinear problem of minimizing regret in the presence of stealthy adversaries as a rank-constrained optimization problem, which can be solved using convex rank minimization methods. We also explored extensions such as incorporating system-level parameterization for sparsity and localizability in online learning, highlighting the potential for predictive, data-driven applications in an interpretable framework.


\bibliographystyle{IEEEtran}
\bibliography{main}

\end{document}